\documentclass[twocolumn]{aastex62}

\usepackage{graphics,graphicx,xspace,natbib,amssymb}
\usepackage[caption=false]{subfig}
\usepackage{amsmath}
\usepackage{threeparttable}
\usepackage{makecell}

\usepackage{multirow}
\usepackage{fontawesome5}

\newcommand{\sersic}{S\'{e}rsic\xspace}

\newcommand{\comment}[1]{}

\newcommand{\fourmic}{4.4\micron\xspace}
\newcommand{\onemic}{1.5\micron\xspace}

\let\itAA\AA
\renewcommand{\AA}{\mathrm{\itAA}}

\shorttitle{Objects in {\it JWST}'s mirror are smaller than they appeared}
\shortauthors{Suess et al.}

\begin{document}

\title{Rest-frame near-infrared sizes of galaxies at cosmic noon:\\objects in {\it JWST}'s mirror are smaller than they appeared}
\shortauthors{Suess et al.}

\author[0000-0002-1714-1905]{Katherine A. Suess}
\affiliation{Department of Astronomy and Astrophysics, University of California, Santa Cruz, 1156 High Street, Santa Cruz, CA 95064 USA}
\affiliation{Kavli Institute for Particle Astrophysics and Cosmology and Department of Physics, Stanford University, Stanford, CA 94305, USA}

\author[0000-0001-5063-8254]{Rachel Bezanson}
\affiliation{Department of Physics and Astronomy and PITT PACC, University of Pittsburgh, Pittsburgh, PA 15260, USA}

\author[0000-0002-7524-374X]{Erica J. Nelson}
\affiliation{Department for Astrophysical and Planetary Science, University of Colorado, Boulder, CO 80309, USA}

\author[0000-0003-4075-7393]{David J. Setton}
\affiliation{Department of Physics and Astronomy and PITT PACC, University of Pittsburgh, Pittsburgh, PA 15260, USA}

\author[0000-0002-0108-4176]{Sedona H. Price}
\affiliation{Max-Planck-Institut f\"{u}r extraterrestrische Physik (MPE), Giessenbachstr. 1, D-85748 Garching, Germany}

\author[0000-0002-8282-9888]{Pieter van Dokkum}
\affiliation{Astronomy Department, Yale University, 52 Hillhouse Ave,
New Haven, CT 06511, USA}

\author[0000-0003-2680-005X]{Gabriel Brammer}
\affiliation{Cosmic Dawn Center (DAWN), Niels Bohr Institute, University of Copenhagen, Jagtvej 128, K\o benhavn N, DK-2200, Denmark}

\author[0000-0002-2057-5376]{Ivo Labb\'{e}} 
\affiliation{Centre for Astrophysics and Supercomputing, Swinburne University of Technology, Melbourne, VIC 3122, Australia}

\author[0000-0001-6755-1315]{Joel Leja}
\affiliation{Department of Astronomy \& Astrophysics, The Pennsylvania
State University, University Park, PA 16802, USA}
\affiliation{Institute for Computational \& Data Sciences, The Pennsylvania State University, University Park, PA, USA}
\affiliation{Institute for Gravitation and the Cosmos, The Pennsylvania State University, University Park, PA 16802, USA}

\author[0000-0001-8367-6265]{Tim B. Miller}
\affiliation{Astronomy Department, Yale University, 52 Hillhouse Ave,
New Haven, CT 06511, USA}

\author[0000-0002-4271-0364]{Brant Robertson}
\affiliation{Department of Astronomy and Astrophysics, University of California,
             Santa Cruz, 1156 High Street, Santa Cruz, CA 95064 USA}

\author[0000-0002-5027-0135]{Arjen van der Wel}
\affil{Astronomical Observatory, Ghent University, Krijgslaan 281, Ghent, Belgium}

\author[0000-0003-1614-196X]{John R. Weaver}
\affil{Department of Astronomy, University of Massachusetts, Amherst, MA 01003, USA}

\author[0000-0001-7160-3632]{Katherine E. Whitaker}
\affil{Department of Astronomy, University of Massachusetts, Amherst, MA 01003, USA}
\affil{Cosmic Dawn Center (DAWN), Denmark}

\email{suess@ucsc.edu}

\begin{abstract}
    Galaxy sizes and their evolution over cosmic time have been studied for decades and serve as key tests of galaxy formation models. However, at $z\gtrsim1$ these studies have been limited by a lack of deep, high-resolution rest-frame infrared imaging that accurately traces stellar mass distributions. Here, we leverage the new capabilities of the {\it James Webb Space Telescope} to measure the \fourmic sizes of ${\sim}1000$ galaxies with $\log{\rm{M}_*/\rm{M}_\odot}\ge9$ and $1.0\le z \le 2.5$ from public CEERS imaging in the EGS deep field. We compare the sizes of galaxies measured from NIRCam imaging at \fourmic ($\lambda_{\mathrm{rest}}\sim1.6\mu $m) with sizes measured at $1.5\mu$m ($\lambda_{\mathrm{rest}}\sim5500\mathrm{\AA}$). We find that, on average, galaxy half-light radii are $\sim9$\% smaller at \fourmic than \onemic\ in this sample. This size difference is markedly stronger at higher stellar masses and redder rest-frame $V-J$ colors: galaxies with ${\rm M}_* \sim 10^{11}\,{\rm M}_\odot$ have \fourmic sizes that are $\sim 30$\,\% smaller than their \onemic sizes. Our results indicate that galaxy mass profiles are significantly more compact than their rest-frame optical light profiles at cosmic noon, and demonstrate that spatial variations in age and attenuation are important, particularly for massive galaxies. The trend we find here impacts our understanding of the size growth and evolution of galaxies, and suggests that previous studies based on rest-frame optical light may not have captured the mass-weighted structural evolution of galaxies. This paper represents a first step towards a new understanding of the morphologies of early massive galaxies enabled by JWST's infrared window into the distant universe.
\end{abstract}


\section{Introduction}

The growth and structural evolution of galaxies over cosmic time provides one of the strongest constraints on theoretical models of galaxy formation, as galaxy sizes are thought to reflect the growth of their host dark matter halos \cite[e.g.,][]{mo98}. However, measuring the sizes of distant galaxies is very difficult from the ground, as atmospheric seeing is roughly the same size as the half-light radii of galaxies beyond $z\sim 0.5$. Characterizing the size growth of galaxies was therefore one of the primary objectives of the \emph{Hubble Space Telescope} ({\it HST}). One of the most exciting and unexpected discoveries from {\it HST} was that quiescent galaxies appeared to grow dramatically with time, more than doubling their sizes between $z\sim2$ until the present day \citep[e.g.,][]{daddi05,vandokkum08,damjanov09,vanderwel14}. This result provided evidence for the importance of gas-poor minor-merging in the growth of massive, quiescent galaxies \citep[e.g.][]{bezanson09,naab09,hopkins09,vandesande13}. {\it HST} also revealed that high-redshift disks grow with time, as expected, but that the growth is slower than expected from basic halo growth models, perhaps implying evolving halo spin parameters \citep[e.g.,][]{somerville08} or varying effects of feedback processes \citep[e.g.,][]{dutton09}. Over a decade of \emph{HST} observations, the literature has coalesced around three basic rules-of-thumb: (1) more massive galaxies tend to be larger; (2) galaxies were smaller at cosmic noon ($z\sim1-2$) than they are in the local Universe; and (3) at fixed mass, star-forming galaxies are larger than their quiescent counterparts \citep[e.g.,][]{bell12,newman12,bruce12,barro13,cassata13,lang14,vanderwel14,vandokkum15,shibuya15,mowla18,nedkova21,cutler22}.

However, these empirical structural measurements are fundamentally limited by the fact that even the longest wavelengths observed by \emph{HST} (1.6$\mu m$) correspond to rest-frame optical light at $z\sim2$, and shift into the rest-frame ultraviolet at higher redshifts. Local galaxies are known to exhibit radial gradients in their mass-to-light (M/L) ratios caused by gradients in their stellar ages, metallicities, or dust attenuation \citep[e.g.,][]{saglia00,labarbera05,tortora10,greene15,woo19,bernardi22}. If these M/L gradients also exist at earlier time, 
then light-weighted size measurements at $z\sim2$ could differ significantly from mass-weighted sizes. This effect is minimized in the rest-frame IR, where K-band light can be used as a reliable proxy for stellar mass; however, at bluer wavelengths the range in M/L varies by up to a factor of ten \citep{bell01}. Given that the stellar populations within galaxies are often highly inhomogeneous, structures measured from rest-frame optical \emph{HST} imaging may be fundamentally biased. These color gradients could be subtle, such as metallicity gradients in an elliptical galaxy, or extreme, such as a composite star-forming galaxy with a large bulge comprised of old stars.
\begin{figure*}[t]
    \centering
    \includegraphics[width=.7\textwidth]{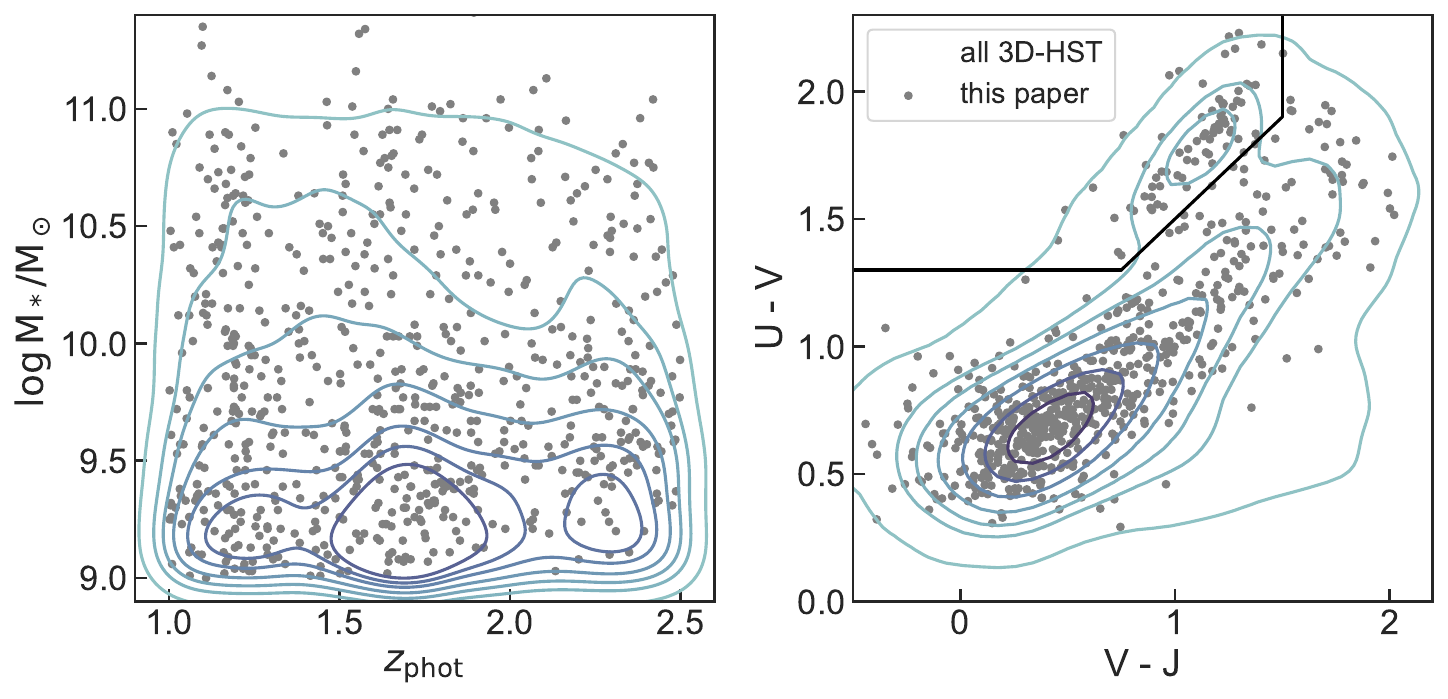}
    \caption{Galaxies in this paper (grey points) compared to all galaxies in the 3D-HST photometric catalog with $\log\rm{M}_*/\rm{M}_\odot>9$ and $1.0\le z\le2.5$ \citep{skelton14}. Our targets follow a similar stellar mass and photometric redshifts distribution as the full 3D-HST catalog, showing that the subset is fairly representative of the full survey. Due to the limited area of the current CEERS observation, relatively few quiescent galaxies (indicated by the box in $UVJ$ color-color space) appear in our sample.}
    \label{fig:selection}
\end{figure*}

Several studies have leveraged color information from multi-band \emph{HST} imaging to approximate mass-weighted structural measurements of distant galaxies \citep{szomoru10,suess19a,suess19b,mosleh20, miller22}. \citet{szomoru13} finds that luminosity-weighted sizes are indeed biased, as they are in the local Universe, but that the effect is small and not strongly dependent on mass or redshift. However, the more recent studies find mass- and redshift-dependent trends, suggesting a less extreme redshift evolution for the structure of massive galaxies. These results call into question just how settled our understanding of galaxy size evolution really is. A more direct understanding of the mass distributions of galaxies requires high-resolution data in the rest-frame infrared: \emph{HST}-based studies must use stellar population models to infer optical M/L and therefore depend on the assumptions that no optically-thick dust is present in the galaxy and that our stellar population models are accurate. Even working at the resolution limit of \emph{HST}'s WFC3 instrument cannot solve these fundamental issues.

The launch and commissioning of the  \emph{James Webb Space Telescope} (\emph{JWST}) enables a vast leap in resolution at infrared wavelengths, for the first time resolving the rest-frame near-infrared structures of galaxies at cosmic noon. Early imaging data in \fourmic from the NIRCam instrument provides nearly mass-weighted light-maps of galaxies at $z\sim1$, in principle enabling the study of the stellar mass distributions of galaxies rather than just their light distributions. In this paper we take a first step in this direction by analyzing images from the Cosmic Evolution Early Release Science (CEERS) program (PI: Finkelstein) to determine whether \fourmic sizes are more compact than those measured from rest-frame optical imaging. 

This Letter is organized as follows: \S2 describes the \emph{HST} and \emph{JWST} data used in this work and our structural measurements. In \S3 we compare \emph{JWST}/NIRCam galaxy sizes at $1.5\mu m$ and $4.4\mu m$. Finally, in \S4 we discuss the implications of these findings and speculate about the exciting structural evolution analysis that \emph{JWST} will enable at cosmic noon and beyond. We adopt a standard $\Lambda {\rm CDM}$ cosmology throughout this paper, with $H_0=70 \; \mathrm{km\,s^{-1}\,Mpc^{-1}}$, $\Omega_m=0.3$, and $\Omega_{\Lambda}=0.7$. All magnitudes are quoted in AB, and we use the term `size' to refer to the major-axis effective radius from a best-fitting S\'ersic model to the galaxy light profile.

\section{Data \& Methods}

The NIRCam F444W and F150W imaging was taken as part of the CEERS program \citep{finkelstein:17} in the AEGIS field. \noindent All the {\it JWST} data used in this paper can be found in MAST: \dataset[10.17909/7v0n-6041]{http://dx.doi.org/10.17909/7v0n-6041}. The F444W observations we use for this project were taken on June 21-28, 2022 with total exposure times of $\sim$1.6$-$6.3 hours per pointing, covering a total of $\sim40$ sq arcmin. We use mosaicked images and weight maps created following the procedure outlined in Brammer et al. (in prep), based off of the public \texttt{grizli} software package \citep{brammer14}.

We use stellar masses and photometric redshift estimates (`z\_best') from the v4.1.5 3D-HST catalog \citep{brammer12,momcheva16,skelton14}. Undoubtedly these masses and redshifts will change for some galaxies in this sample given the new long-wavelength photometry now available in these fields; however, for this first-look paper our primary goal is to compare the sizes of these galaxies measured at \onemic and \fourmic. At the most basic level, this test can be performed in on-sky units and is independent of stellar population fitting parameters.

Our sample consists of all galaxies at cosmic noon in the CEERS field that are both massive and sufficiently bright to allow straightforward structural measurements, corresponding to a selection of $1.0\le z\le2.5$, $\log{\rm{M}_*/\rm{M_\odot}}\ge9$, `use\_phot'=1 from the 3D-HST catalogs, and coverage in the initial CEERS imaging mosaic. Following \citet{vanderwel12}, we use an HST/F160W magnitude cut of 24.5 to ensure robust fits. These cuts result in a sample of 1179 galaxies. The `use\_phot' flag is to indicate photometry of reasonable quality, avoiding e.g. stellar diffraction spikes and CCD defects, and thus does not bias the sample \citep{skelton14}. 
\begin{figure*}
    \centering
    \includegraphics[width=.95\textwidth]{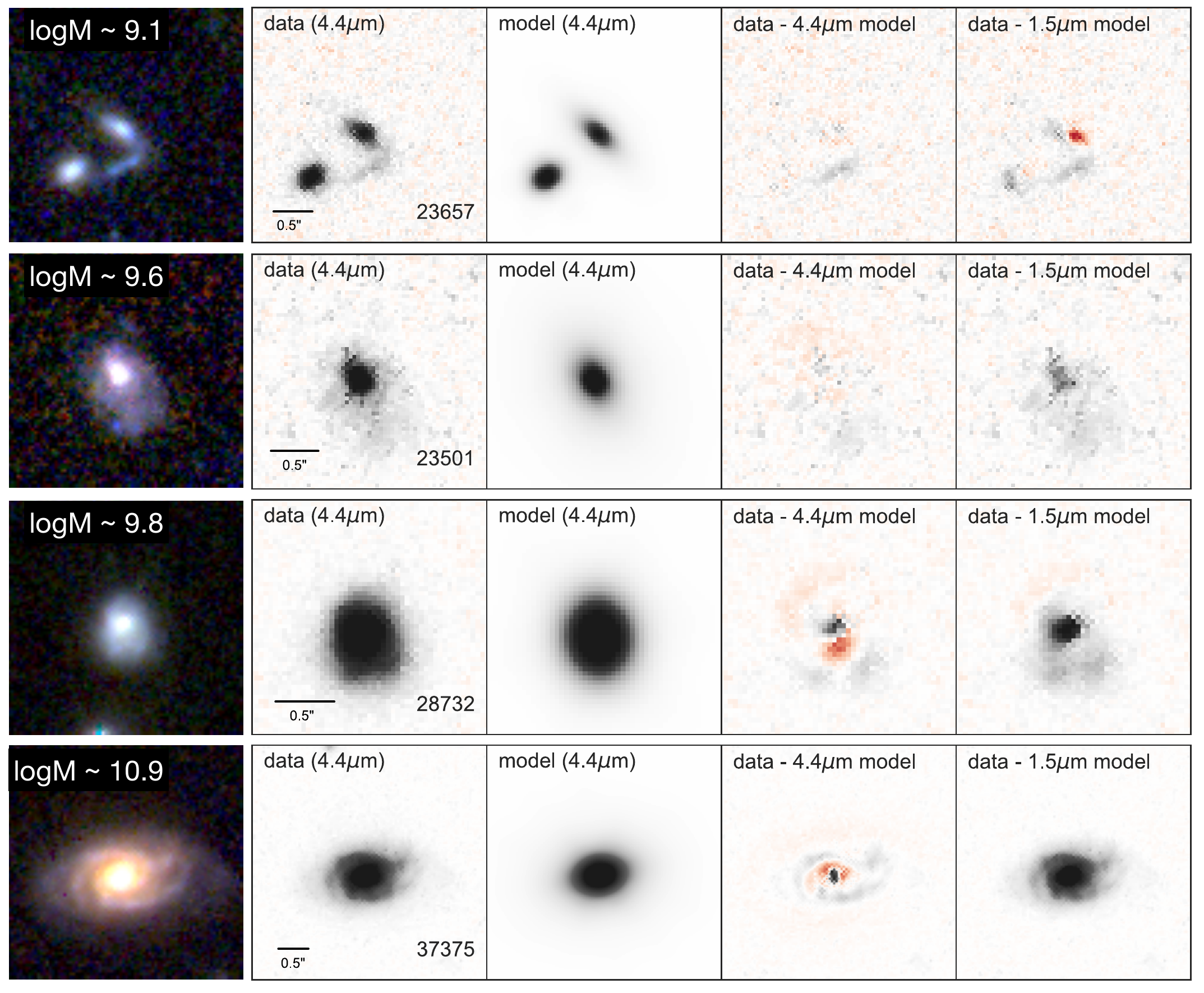}
    \caption{Example \texttt{GALFIT} fits for four objects in our sample, ordered from lowest to highest stellar mass. From top to bottom, the objects lie at redshifts of 1.9, 2.2, 1.2, and 1.5. The left column shows three-color images using \emph{JWST}/NIRCam F150W, F277W, and F444W. The center three panels show the 4.4$\mu$m data, our best-fit model and residual. For comparison, the right panel shows the difference between the \fourmic data and the best-fit 1.6$\mu$m \texttt{GALFIT} model convolved with the 4.4$\mu$m PSF. The data, model, and residuals are on the same symmetric colorbar for each object, with positive values in greyscale and negative values in red. Our fitting procedure is able to accurately reproduce galaxy cutouts for both isolated and crowded fields, but cannot capture complex morphologies such as spiral arms. The 1.6$\mu$m model tends to underpredict the flux at the center of the galaxy; this is especially true for the higher-mass galaxies in the bottom rows of the plot.}
    \label{fig:example_fits}
\end{figure*}

We fit the sizes of all 1179 galaxies in our sample using the \texttt{GALFIT} software package \citep{peng02}. As inputs, \texttt{GALFIT} requires an image, a weight map, and a point spread function (PSF). Because the centers of most stars are saturated in our CEERS mosaic, creation of an empirical PSF is impractical. We therefore use theoretical PSFs generated using the WebbPSF software \citep{perrin14}. By default, WebbPSF generates PSFs at a position angle of zero. Rotating these PSFs to match the position angle of the observations could introduce distortions unless the rotation is performed on an oversampled PSF. Therefore, we use WebbPSF to generate 9x-oversampled F444W and F150W PSFs assuming the same 0.04" pixel scale of our mosaicked image. We then rotate the oversampled PSFs to the position angle of the CEERS exposures, convolve with a 9x9 square kernel, and downsample back to the 0.04" mosaic pixel scale.

\begin{figure}
    \centering
    \includegraphics[width=.45\textwidth]{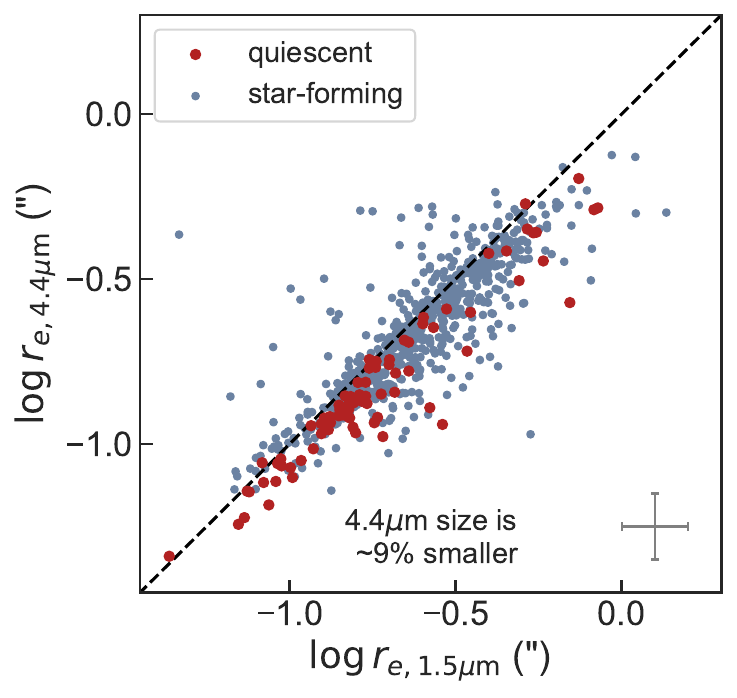}
    \caption{\fourmic size as a function of \onemic size in arcseconds, colored by star-forming or quiescent based on rest-frame $UVJ$ colors. Galaxies are slightly smaller in \fourmic than in \onemic, the longest-wavelength filter previously accessible with \emph{HST}. The lower right shows a typical error bar estimated from comparisons to \citet{vanderwel14} (see text).}
    \label{fig:size_comparison}
\end{figure}

We choose an F444W cutout size of $100\times r_{e,\rm{F160W}}$ as measured in the \citet{vanderwel12} HST-based size catalog, enforcing a minimum cutout size of 80x80 pixels and a maximum cutout size of 200x200 pixels and adopting a size of 150x150 pixels for galaxies without existing HST structural measurements. We use standard \texttt{astropy} and \texttt{photutils} procedures to create a segmentation map of each cutout to identify any additional galaxies in the image. We simultaneously model any galaxies that have magnitudes up to 2.5~mag fainter than the target galaxy and have centers within 2" of the center of the cutout. We mask fainter or more distant galaxies. After performing segmentation, we estimate and subtract off a scalar local background correction using the \texttt{photutils} implementation of the SExtractor background subtraction scheme. Finally, we fit each background-subtracted cutout with \texttt{GALFIT} to determine the \fourmic size of these {\it HST}-selected galaxies. We repeat this procedure with F150W cutouts to measure $150\mu$m sizes. We remove from our sample 400 galaxies where \texttt{GALFIT} crashed or returned bad flags --- typically, this occurs because one or more of the fit parameters reached our bounds on size or \sersic index. We additionally flag and remove 15 galaxies where the best-fit \texttt{GALFIT} \onemic or \fourmic magnitude differs by $>2$~mag from the magnitude as measured in our CEERS catalog. This offset typically occurs when our pipeline mistakenly fits a nearby bright galaxy instead of the fainter target galaxy. In total, we provide robust sizes for 703 of the 1179 of the galaxies in our parent sample. The fraction of galaxies successfully fit by our pipeline is slightly higher ($\sim$60\%) than the fraction of galaxies with no flags in the  \citet{vanderwel12} \emph{HST} size catalog ($\sim$50\%).

We test the robustness of our measurement pipeline by comparing our measured \onemic sizes to the $1.6\mu$m sizes measured by \citet{vanderwel14}. While the data depth, PSF size, and exact fitting details differ between these two sets of measurements, the underlying sizes should be relatively similar given the small difference in wavelength. We find that our \onemic sizes are not systematically biased compared to the \citet{vanderwel14} catalog, with an offset of $<0.01$~dex and a scatter of $\sim0.12$ dex. Given the significant differences between the two datasets and the analysis pipeline (e.g., source masking and background subtraction schemes), we believe this agreement is very robust. Based on this test, we adopt a systematic uncertainty on our size measurements, included as a characteristic errorbar in Fig. 3. 

Table~\ref{tab:results} provides our morphological measurements at both \onemic and \fourmic. ID numbers correspond to the v4.1 3D-HST catalog \citep{skelton14}. Error bars listed are from \texttt{GALFIT} and likely underestimated; systematic error bars on sizes are estimated to be $\sim0.15$~dex from mock recovery tests. We do not include integrated magnitudes in Table~\ref{tab:results} as these are subject to still-evolving knowledge of exact photometric zeropoints; our structural parameters and main results are not sensitive to any zeropoint offsets. 

\begin{table*}
\caption{Morphological measurements at \onemic and \fourmic.}
\footnotesize
\begin{tabular}{ccccccccccc}
\hline \hline
ID & log$M_{\star}$ & z & $r_e$ (\onemic) & $n$ (\onemic)  & $q$ (\onemic)  & PA (\onemic)  & $r_e$ (\fourmic)  & $n$ (\fourmic)  & $q$ (\fourmic)  & PA (\fourmic) \\
\hline
13225 & 9.31 & 2.48 & $0.56 \pm 0.01$ & $2.04 \pm 0.07$ & $0.48 \pm 0.01$ & $-11.14 \pm 0.64$ & $0.67 \pm 0.01$ & $1.84 \pm 0.11$ & $0.55 \pm 0.01$ & $-14.80 \pm 1.14$ \\
13428 & 9.84 & 1.47 & $2.22 \pm 0.03$ & $0.91 \pm 0.02$ & $0.65 \pm 0.01$ & $14.34 \pm 0.86$ & $2.11 \pm 0.01$ & $0.73 \pm 0.01$ & $0.57 \pm 0.00$ & $24.23 \pm 0.34$ \\
13486 & 9.70 & 1.27 & $3.68 \pm 0.03$ & $1.07 \pm 0.01$ & $0.29 \pm 0.00$ & $-24.58 \pm 0.15$ & $3.28 \pm 0.01$ & $1.05 \pm 0.01$ & $0.31 \pm 0.00$ & $-23.57 \pm 0.11$ \\
13567 & 9.08 & 1.49 & $2.30 \pm 0.02$ & $0.44 \pm 0.01$ & $0.27 \pm 0.00$ & $-55.88 \pm 0.23$ & $2.29 \pm 0.01$ & $0.39 \pm 0.01$ & $0.31 \pm 0.00$ & $-56.20 \pm 0.24$ \\
13598 & 9.86 & 1.27 & $4.39 \pm 0.04$ & $0.94 \pm 0.01$ & $0.54 \pm 0.00$ & $25.63 \pm 0.40$ & $3.04 \pm 0.01$ & $1.65 \pm 0.01$ & $0.45 \pm 0.00$ & $20.53 \pm 0.12$ \\
13626 & 9.71 & 1.19 & $1.11 \pm 0.02$ & $2.90 \pm 0.06$ & $0.80 \pm 0.01$ & $-52.96 \pm 1.42$ & $0.60 \pm 0.00$ & $2.64 \pm 0.04$ & $0.75 \pm 0.00$ & $-59.93 \pm 0.69$ \\
13647 & 9.05 & 1.44 & $2.37 \pm 0.05$ & $1.20 \pm 0.03$ & $0.22 \pm 0.00$ & $-10.21 \pm 0.32$ & $2.43 \pm 0.02$ & $0.71 \pm 0.02$ & $0.25 \pm 0.00$ & $-5.99 \pm 0.23$ \\
13795 & 10.09 & 1.28 & $2.09 \pm 0.02$ & $1.60 \pm 0.02$ & $0.61 \pm 0.00$ & $-44.34 \pm 0.43$ & $1.63 \pm 0.00$ & $1.66 \pm 0.01$ & $0.61 \pm 0.00$ & $-47.61 \pm 0.20$ \\
13842 & 10.36 & 1.47 & $2.81 \pm 0.06$ & $2.89 \pm 0.05$ & $0.86 \pm 0.01$ & $-26.65 \pm 2.19$ & $1.25 \pm 0.01$ & $1.72 \pm 0.03$ & $0.79 \pm 0.01$ & $-6.67 \pm 1.03$ \\
13908 & 9.74 & 1.83 & $2.45 \pm 0.02$ & $0.53 \pm 0.01$ & $0.30 \pm 0.00$ & $20.52 \pm 0.21$ & $1.99 \pm 0.01$ & $0.79 \pm 0.01$ & $0.36 \pm 0.00$ & $18.97 \pm 0.19$ \\
... & ... & ... & ... & ... & ... & ... & ... & ... & ... & ... \\
\end{tabular}
\begin{tablenotes}
\item (This table is available in its entirety in a machine-readable form in the online journal. A portion is shown here for guidance regarding its form and content.)
\end{tablenotes}
\label{tab:results}
\end{table*}

Figure~\ref{fig:example_fits} shows several example fits in \fourmic. The left three columns show the \fourmic data cutout, model, and residual. The rightmost column highlights the differences in structures between the \fourmic and \onemic imaging. We use \texttt{GALFIT} to generate a \sersic model with the best-fit parameters of the \onemic model, but convolve it with the \fourmic PSF. The figure shows the difference between the \fourmic data and this \onemic fit. In all cases, we see residual flux remaining at the center of the galaxy that is not accounted for by the \onemic model, indicating that \fourmic sizes are smaller than \onemic sizes.

We note several caveats before moving on to discuss our results. First, we use a theoretical PSF rather than an empirical one. While this choice is necessary because the centers of stars are clipped/saturated in our mosaic, a theoretical PSF could introduce errors into the estimation of intrinsic galaxy shapes because the theoretical PSF is not a perfect representation of the true PSF in the images. We also do not account for any variations in the PSF across the field of view. Second, while we have subtracted off our best estimate of a scalar sky background, the backgrounds in this early NIRCam imaging retain some structure which may impact the modeling of the wings of the light distribution in galaxies. Third, as can be seen in Fig.\ref{fig:example_fits}, many galaxies are not well approximated by the simple sersic profile with which we model them. Many exhibit e.g. clumps and spiral arms which are not accounted for in our model and hence affect our parametric fit. One way to solve this in the future may be to move towards nonparametric models \citep[e.g.][]{miller:22}. 

\section{Results}
In Figure~\ref{fig:size_comparison}, we compare \emph{JWST} \fourmic\ (F444W) sizes to \onemic (F150W) sizes, measured in arcseconds. These \onemic sizes trace similar stellar populations as previous size measurements in \emph{HST}/F160W \citep[e.g.,][]{vanderwel12}. 
We stress that this size comparison does {\it not} depend on stellar population modeling or photometric redshift estimates for these galaxies: we are simply comparing the on-sky extent of each galaxy. Points are colored by their designation as either star-forming or quiescent based on their rest-frame $UVJ$ colors as measured in the 3D-HST catalog, using the same quiescent definition as \citet{whitaker12}. 
Figure~\ref{fig:size_comparison} shows that galaxies are systematically smaller in \fourmic\ than they appeared in \onemic. Galaxies are 9\% smaller on average at \fourmic\ than they appear in \onemic. This systematic offset is a factor of $\sim4$ larger than the difference between our \onemic sizes and the \citet{vanderwel12} sizes, and is unlikely to be due to measurement error. 
With our newfound capacity to resolve the rest-frame near-infrared emission in galaxies at cosmic noon with \emph{JWST}, we find that galaxies are more compact than they appeared when we could only observe their rest-frame optical emission with \emph{HST}. 

\begin{figure*}
    \centering
    \includegraphics[width=.95\textwidth]{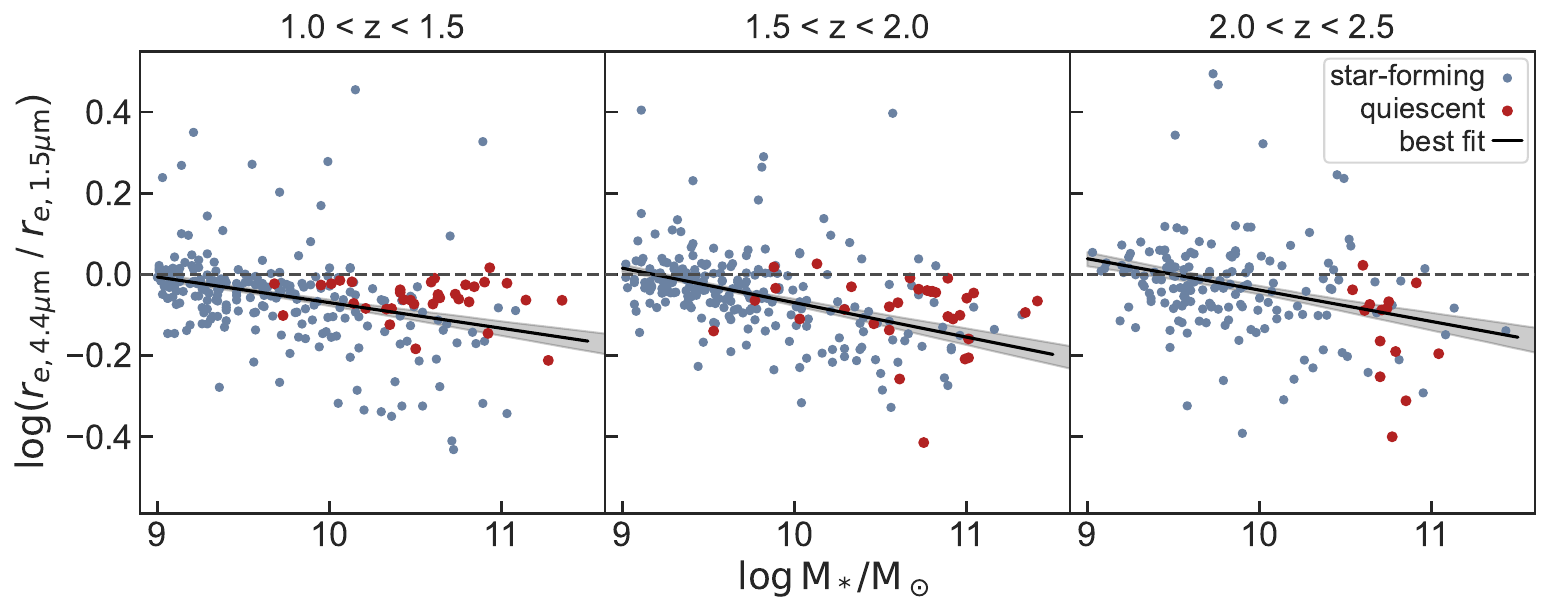}
    \includegraphics[width=.95\textwidth]{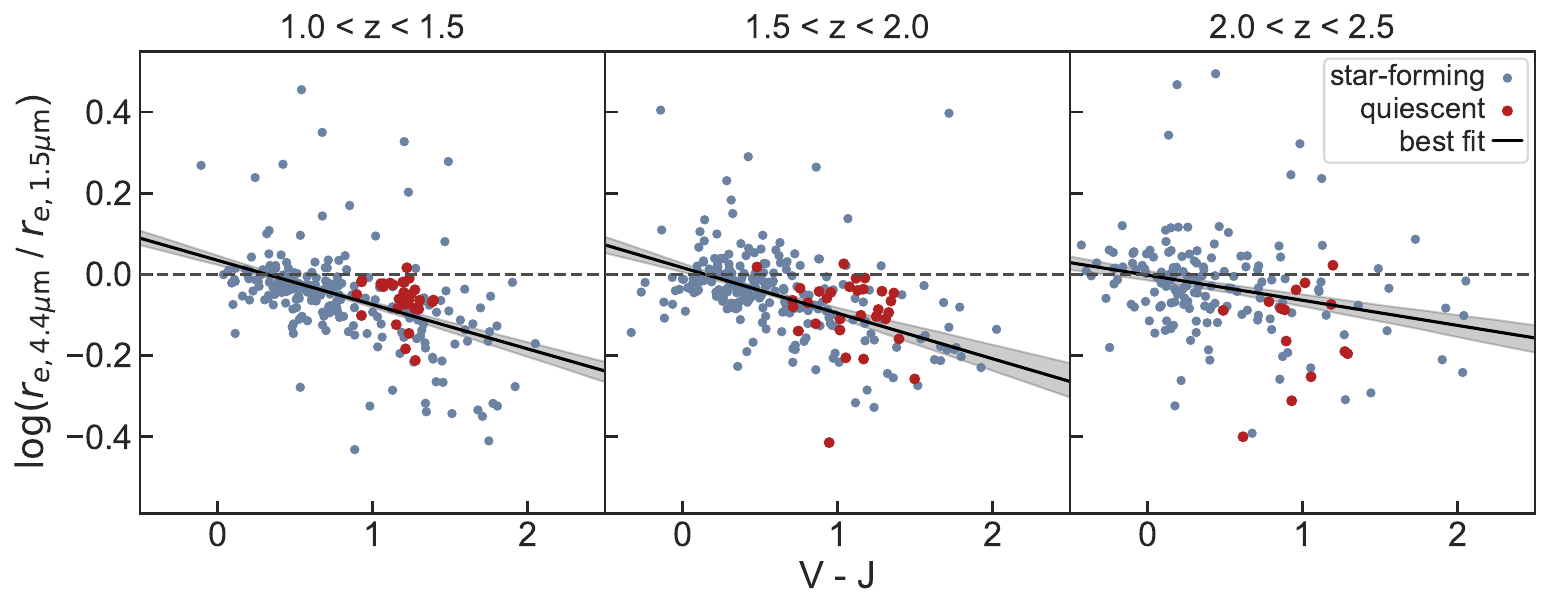}
    \caption{Estimate of color gradient strength as probed by the ratio of \fourmic to \onemic sizes as a function of both stellar mass and rest-frame $V-J$ color in three redshift bins. Marker color indicates $UVJ$-selected star-forming and quiescent galaxies. We find significant trends with both mass and color, such that redder, more massive galaxies have larger differences between their \onemic and \fourmic sizes. These color gradients indicate that more massive galaxies have increasingly redder centers compared to their outskirts.}
    \label{fig:cgrad_mass}
\end{figure*}

In Figure~\ref{fig:cgrad_mass}, we explore how the ratio of observed \fourmic size to \onemic size changes with stellar mass and rest-frame $V-J$ color. Each panel shows a different redshift slice, because the \fourmic and \onemic filters probe different rest-frame wavelengths in each bin, with rest-frame wavelength ranging between $\sim1.3-2\mu$m for \fourmic and $\sim4600-6600\AA$ for \onemic. In all three redshift bins, we see a trend such that redder and more massive galaxies have stronger color gradients. While galaxies are color-coded according to their star-formation status, due to the limited area of our current study we do not have sufficient numbers of quiescent galaxies to fit the star-forming and quiescent trends separately. We perform a linear fit to the logarithmic size ratio for the entire population using \texttt{scipy}'s `curve\_fit' function, estimating uncertainties in the best-fit parameters using bootstrapping. We find that the slope of the $\log(r_{e,\rm{\fourmic}} / r_{e,\rm{\onemic}}) - \log{\rm{M}_*/\rm{M}_\odot}$ trend is significant,  $\sim-0.05\pm0.02$ in all three redshift bins. This mass effect is strong: at $10^9\rm{M}_\odot$, we find that galaxies are almost exactly the same size in \fourmic and \onemic; by $10^{11}\rm{M}_\odot$, galaxy \fourmic sizes are just $\sim70$\% of their \onemic sizes. 

\section{Discussion \& Conclusions}

In this Letter, we investigate the effect that new information from {\em JWST} --- diffraction-limited space-based \fourmic imaging --- has on the measured sizes of galaxies at cosmic noon. While the size-mass relation was thought to be relatively well-understood over this mass and redshift regime \citep[e.g.,][]{vanderwel14,mowla18,kawinwanichakij21}, recent work has called into question whether this view is biased by the rest-optical nature of \emph{HST} observations  \citep[e.g.,][]{suess19a,suess19b,suess20,mosleh20,miller22}. 
These new \emph{JWST} observations allow us to directly measure the rest-frame infrared morphologies of galaxies --- a reasonably reliable proxy for their stellar mass distribution --- at cosmic noon for the first time, testing whether our previous picture of size evolution was biased by the available data. 

We find that galaxies tend to be smaller at \fourmic than they are at $1.5\mu$m (Figure~\ref{fig:size_comparison}). This mirrors results in the local universe \citep[e.g.,][]{kelvin12,lange15}, and shows that galaxy structures were already complex
8--10\,Gyr ago.
Strikingly, the effect strongly depends on galaxy properties: more massive galaxies and galaxies with redder $V-J$ colors have stronger color gradients than their less massive, bluer counterparts. On average, $10^9\rm{M}_\odot$ galaxies have the same \fourmic and \onemic sizes; by $10^{11}\rm{M}_\odot$, \fourmic galaxy sizes are $\sim 30$\,\% smaller than their $1.5\mu$m sizes. These results qualitatively agree with previous studies which have used stellar population synthesis modelling to measure galaxy mass profiles: these studies have generally found that galaxy half-mass radii are smaller than half-light radii, with the effect becoming more important at higher stellar masses \citep{suess19a,miller22}. However, we emphasize the novel, model-independent nature of this result; all previous studies have relied on stellar population synthesis modeling to infer M/L ratios. This modeling is known to be sensitive to a host of systematic uncertainties due to e.g., dust-metallicity-age degeneracies and even the initial mass function. The fact that galaxies do indeed appear smaller at 4\micron\ points to an exciting new era of stellar mass-weighted structural studies with \emph{JWST}.

The differences we observe between $1.5\mu$m and \fourmic sizes call into question our understanding of the galaxy size-mass relation. Because more massive galaxies tend to have stronger \fourmic/$1.5\mu$m size differences, the inferred slope of the size-mass relation flattens: more massive galaxies may {\it not} be larger than less massive ones. 
If the strength of color gradients evolves with redshift \citep[as suggested by][]{suess19a,suess19b}, then galaxies at cosmic noon may {\it not} be significantly smaller than their local counterparts. If color gradients differ between star-forming and quiescent galaxies --- beyond the scope of our current limited sample size, but perhaps suggested by the trends with color we observe in Figure~\ref{fig:cgrad_mass} --- then quiescent galaxies may {\it not} be smaller at fixed mass than star-forming galaxies. All three of our \emph{HST}-based ``rules of thumb" about galaxy sizes may change with \emph{JWST}'s new window into the infrared universe. While this first-look study focused only on the sizes of bright, \emph{HST}-selected galaxies, our picture of the galaxy size-mass relation will undoubtedly change with \emph{JWST}.

The implications of our observed size differences go beyond mitigating biases in our understanding of galaxy sizes: they allow us to understand {\it how} galaxies assemble their stellar mass. Differences in morphology at different rest-frame wavelengths can be mapped back to physical quantities --- primarily radial variations in age and dust, but also of stellar metallicities. This means that the size differences seen in Figures~\ref{fig:size_comparison} and \ref{fig:cgrad_mass} can be used to understand variations in stellar population properties of galaxies. Smaller \fourmic sizes than \onemic sizes imply that stellar mass profiles are more compact than light profiles, indicating redder centers. These redder centers may be due to dust --- previous work at cosmic noon has showed that galaxy centers tend to be more dust-obscured than their outskirts \citep[e.g.][]{nelson16a}, that more massive galaxies tend to be dustier \citep[e.g.,][]{whitaker17}, and that galaxies with redder $V-J$ colors are more likely to be edge-on disks with very obscured centers \citep[e.g.,][]{patel12}. Or, these redder centers may be due to older stellar ages --- e.g., these massive disk galaxies may be in the process of assembling the bulge components that we see in massive galaxies in the local universe \citep[e.g.][]{bezanson:09,bezanson:11,nelson:19,tadaki:20}. Our finding that even half-light radii, one the most basic measure we have of galaxy morphologies, differ by up to $\sim30$\% between \fourmic and \onemic in massive galaxies indicates that our previous understanding of the true structures of massive galaxies at cosmic noon was incomplete. Although these differences may seem subtle, they could prove fundamental to our understanding of how galaxies quench and structurally transform. If light-weighted size estimates are indeed systematically biased and star-forming galaxies are closer in size to their quiescent counterparts at cosmic noon, this could alleviate the need for dramatic structural transformation \citep[e.g.,][]{zolotov15} or careful progenitor and descendent matching \citep[e.g.][]{vandokkum15}.

Moving forward, the remarkable public datasets gathered with \emph{JWST} can be used to study {\it why} we see these differences in galaxy sizes across wavelengths. To date, quantitative measurements of age, dust, and metallicity gradients in galaxies have primarily been restricted to the local Universe \citep[e.g.,][]{greene12,greene15,woo19}. New \emph{JWST} multi-band infrared imaging, along with a legacy of UV-optical imaging from \emph{HST}, will allow us to extend spatially-resolved stellar population fitting methodologies to quantify age and dust gradients in galaxies at cosmic noon, and to place strong constraints on their underlying stellar mass distributions. Longer wavelength data with \emph{JWST}/MIRI or ALMA may additionally help to constrain the sizes of massive star-forming galaxies at cosmic noon \citep[e.g.,][]{franco:20,valentino:20,Gomez-Guijarro:22}.
These measurements can be directly compared to predictions from simulations \citep[e.g.,][]{x_wu20,pathak21,marshall22} in order to gain a more complete picture of galaxy growth and assembly.

\acknowledgements 
KAS acknowledges the UCSC Chancellor’s Postdoctoral Fellowship Program for support. BER acknowledges the use of the lux supercomputer at UC Santa Cruz, funded by NSF MRI grant AST 1828315, and support from NASA grants 80NSSC18K0563, 80NSSC22K0814, and the NIRCam science team.  The Cosmic Dawn Center (DAWN) is funded by the Danish National Research Foundation under grant No. 140. Cloud-based data processing and file storage for this work is provided by the AWS Cloud Credits for Research program. EJN acknowledges support from HST-AR-16146. RB acknowledges support from the Research Corporation for Scientific Advancement (RCSA) Cottrell Scholar Award ID No: 27587.

\software{astropy \citep{astropy2013, astropy2018},  
          GALFIT \citep{peng02}, grizli \citep{brammer14}, Seaborn \citep{waskom17}, WebbPSF \citep{perrin14}
          }

\bibliographystyle{aasjournal}
\bibliography{all}

\end{document}